\documentclass[twocolumn,showpacs,aps,superscriptaddress,amsmath,amssymb]{revtex4}
\usepackage{graphicx}
\usepackage{amsmath}
\usepackage{amssymb}
\usepackage{array}
\begin{document}
\title{Entanglement purification for high-dimensional multipartite systems}

\author{Yong Wook Cheong}

\affiliation{Quantum Photonic Science Research Center, Hanyang
  University, Seoul 133-791, Korea}

\author{Seung-Woo Lee}

\affiliation{Clarendon Laboratory, University of Oxford, Parks Road,
  Oxford OX1 3PU, United Kingdom}

\author{Jinhyoung Lee}

\affiliation{Quantum Photonic Science Research Center, Hanyang
  University, Seoul 133-791, Korea}

\affiliation{BK21 Program Division of Advanced Research and Education in Physics, Hanyang University, Seoul 133-791, Korea}

\affiliation{Department of Physics, Hanyang University, Seoul 133-791,
  Korea}

\author{Hai-Woong Lee}

\affiliation{Department of Physics, Korea Advanced Institute of Science
  and Technology, Daejeon 305-701, Korea}

\date{\today}

\begin{abstract}
  We propose an entanglement purification protocol for high-dimensional multipartite systems. In the protocol we can select a subensemble in a pure generalized Greenberger-Horne-Zeilinger (GHZ) state. This post-selection can be made by detecting the noise which contaminated the initial pure ensemble when the systems past through a noisy environment. For the detection of noise we investigate correlation properties of GHZ states and analyze their possible errors due to a noisy environment. We show that the presented protocol is more efficient than a simple generalization of the purification protocol for a bipartite state in high dimensions.
\end{abstract}

\pacs{03.67.Mn,03.67.Pp,03.67.Hk}

\maketitle

\newcommand{\bra}[1]{\left<#1\right|}
\newcommand{\ket}[1]{\left|#1\right>}
\newcommand{\abs}[1]{\left|#1\right|}

\section{introduction}
\label{sec:int}

Entanglement lies at the heart of quantum information science. For instance, quantum entangled channels are essential in teleporting an unknown quantum state \cite{bennett01} and in sharing a secret key for cryptography \cite{ekert01}. On the other hand, entanglement displays very fragile behavior in a noisy environment: When a system initially in a pure entangled state comes into contact with an environment, the state will decohere to a mixed state as it is unavoidably entangled with the environment and eventually will totally lose its entanglement \cite{kim02}. This decoherence will decrease the degree of entanglement in a quantum channel for teleporation and degrade its fidelity \cite{lee00}. Similarly, the decoherence will bring insecurity to the encrypted communication. Entanglement purification protocols (EPP's) have been developed to recover a subensemble in a maximally entangled pure state from a decohered ensemble before performing reliable quantum information processes \cite{bennett02,bennett03,murao,dur,chen,horo,delgado,duan,miyake,nest,pan}.
They have been applied to the distillation of Bell states of qubits \cite{bennett02,bennett03}
and many-qubit entangled states \cite{murao,dur,chen,miyake}. An experimental realization has also been reported \cite{pan}. Recently, extensions of EPP's have been proposed for bipartite high dimensional systems and continuous variables \cite{horo,delgado,duan}.

Multipartite entangled states have been investigated intensively for quantum information processes, mostly based on the use of qubits. These studies include Greenberger-Horne-Zeilinger (GHZ) nonlocality \cite{ghz}, quantum secret sharing \cite{buzek}, cluster-state computation \cite{rauss}, and quantum error correcting codes (QECC's) \cite{shor,gottesman,knill,braun}. Recently, strongly coupled systems have attracted interest due to their potential for scalable quantum computation and quantum simulation \cite{greentree06}. States of these systems are defined on high dimensional Hilbert spaces and are typically multipartite $d$-dimensional systems, also known as qudits. It is desirable to understand what characteristics they have in terms of entanglement, what information processes they can be made to perform and if any advantages can be obtained over qubits. It has been shown that qudits enable more secure quantum cryptography than qubits \cite{gisin, lee02}. Two entangled qudits exhibit increased robustness against isotropic noises in nonlocality \cite{cglmp02}. Studies for many qudit systems include nonlocality \cite{lee}, entanglement swapping \cite{bouda}, and preparation of GHZ states using hyper entanglement \cite{Ren}. Optical realization of entanged qudits has recently been reported by using photon polarization \cite{howell}, orbital angular momentum \cite{Mair}, and spatial modes \cite{Neves}. It was also reported that a sequence of pump pulses leads to time-bin entanglement of qudits \cite{riedmantten}.

Entanglement purification is concerned with how to obtain a reduced number of copies of a high-fidelity entangled state from many polluted copies of noisy/nonmaximally entangled states, while quantum error correction focuses on how to protect a certain quantum state from noise by encoding the state in a larger dimensional Hilbert space. Both approaches share a common purpose of restoring the quantum coherence and they are closely related \cite{dur}. Indeed, a QECC can be constructed from a one-way EPP, where only uni-directional classical communication is allowed, and vice versa \cite{bennett03}. This fact was used for proving the security of quantum key distribution \cite{shor02}. EPP's may be classified into recursive, hash, and breed methods \cite{bennett03}. Recursive two-way EPP's with bi-directional classical communication \cite{bennett02,bennett03,murao,dur,chen,horo} can be constructed and analyzed in the frame of quantum error detecting codes (QEDC's) \cite{gottesman02,matsumoto,vaid}, while other EPP's utilizing local permutations are also possible \cite{nest}.

In this paper, we propose a recursive two-way EPP for GHZ entangled states of many qudits. For the purpose we analyze correlation properties of generalized GHZ states
(which will be called ``GHZ states'' throughout the paper). Based on the analysis, we find a measurement procedure to detect possible errors in the contaminated GHZ state as transmitted through a noisy environment.  Then, we present an EPP that can be employed for purifing a many-qudit GHZ state.  The presented protocol is shown to be more efficient than a simple generalization of modifying the protocol suggested previously to purify a two-qudit entangled state \cite{horo}.

\section{Perfect Correlations for GHZ States}
\label{sec:cor}

Before considering the purification scheme, we first discuss correlation properties of GHZ states. A maximal GHZ state shows perfect correlation properties for certain composite observables and based on them it can be used for quantum communication. As subsystems in a maximal GHZ state are distributed through a noisy environment, their interaction with the environment can contaminate the GHZ state with some errors and simultaneously alters correlation properties in the GHZ state. By examining the alteration, we can reciprocally detect the errors caused by the noisy environment. Based on the detection result, we select copies with no errors. This post-selection increases the fidelity to the maximal GHZ state. Thus, investigating correlation properties of GHZ states are of importance in a purification scheme.

Suppose that three observers, Alice, Bob, and Charlie want to share three qudits for quantum communication and further that the composite system of three qudits are in one of the GHZ states, defined by
\begin{equation}
\label{eq:ghzs}
 |\psi_{lmn} \rangle_{ABC} =\frac{1}{\sqrt{d}} \sum_{k=0}^{d-1}
 \omega^{lk}|k,k\ominus m,k \ominus n
 \rangle,
\end{equation}
where $l,m,n \in \{ 0,1,\ldots,d-1\}$, $\omega=e^{i2\pi/d}$, $\{|k\rangle\}$ is an orthonormal basis, and $k\ominus m$ denotes $k-m \mod d$, {\em i.e.}, the residue when $k-m$ is divided by modulus $d$. Here $d$ is the dimension of Hilbert space $H_d$ for a qudit and the subscripts $A$, $B$, and $C$ stand for Alice, Bob, and Charlie, respectively. Note that the tensor product symbol ``$\otimes$''  is omitted for a sake of simplicity in Eq.~(\ref{eq:ghzs}). These GHZ states form a complete basis set in the Hilbert space $H_d \otimes H_d \otimes H_d$ of the composite system.

Each observer is allowed to choose one of two observables, $x$ and $z$. An observable is assumed to take, as its value, an element in the set $S=\{1, \omega, \dots, \omega^{d-1}\}$.  We introduce operators $X$ and $Z$ to represent orthogonal measurements for the observables $x$ and $z$. An orthogonal measurement is represented by a complete set of orthonormal eigenvectors and its outcomes are denoted by a set of eigenvalues. As the observables $x$ and $z$ are assumed to take an element in $S$, let $S$ be the set of eigenvalues and then the observable operators are given by
\begin{eqnarray}
\label{eq:xz}
X &=& \sum_{n=0}^{d-1} \omega^k |k\rangle_{xx} \langle k|, \nonumber \\
Z &=& \sum_{n=0}^{d-1} \omega^k |k\rangle_{zz} \langle k|,
\end{eqnarray}
where $|k\rangle_{x,z}$ are eigenvectors of $X$ and $Z$, respectively, corresponding to an eigenvalue $\omega^k$. The integer $k$ of $|k\rangle$ will be called an outcome throughout the paper. In this representation the observable operators $X$ and $Z$ are unitary \cite{lee}. To be more specific, let the measurement basis of $Z$ be a standard basis, that is,  $\{|k\rangle_z=|k\rangle\}$. The measurement basis of $X$ is defined by quantum Fourier transformation $F$ on $\{|k\rangle_z\}$, {\em i.e.}, each eigenvector of $X$ is given by
\begin{eqnarray}
  \label{eq:loofx}
  |k\rangle_x = F |k\rangle_z = \frac{1}{\sqrt{d}}\sum_{j=0}^{d-1}
   \omega^{k j}|j\rangle
\end{eqnarray}
for an eigenvalue $\omega^k$. Then, the observable operator $X$ is a level-shift operator in the standard basis as
\begin{eqnarray}
  \label{eq:ooxisb}
  X=\sum_{k=0}^{d-1} \ket{k} \bra{k+1 \mod d}.
\end{eqnarray}

To describe a correlation, we employ a conditional probability among the outcomes $p$, $q$, and $r$ that three observers obtain from their local measurements, respectively.
Suppose Bob and Charlie try to infer Alice's outcome $p$, based on their obtained outcomes $q$ and $r$. The conditional probability of their inference being correct is
given by
\begin{equation}
\label{eq:cpft}
P(p|q,r)=\frac{P(p,q,r)}{P(q,r)},
\end{equation}
where $P(p,q,r)$ is a joint probability of Alice, Bob, and Charlie obtaining outcomes $p$, $q$, and $r$, respectively and $P(q,r)$ is that of Bob and Charlie obtaining outcomes $q$ and $r$. If $P(p|q,r) \neq P(p)$, Alice's outcome is correlated to those of Bob and Charlie. The correlation is said to be perfect when a pair of observers, based on their outcomes, can infer with a certainty the other's outcome: $P(p|q,r)=1$ for certain $p$, $q$, and $r$.

To investigate correlations for a given GHZ state $\ket{\psi_{lmn}}$ in Eq.~(\ref{eq:ghzs}), we need joint probabilities and conditional probabilities for composite measurements. A composite measurement is denoted as $x_A x_B x_C$
if three observers perform all $X$ measurements on their received qudits, respectively. Similarly we denote another as $z_A z_B$ if Alice and Bob perform $Z$ measurements with Charlie performing no measurement. From the measurement $x_A x_B x_C$, the observers have joint probabilities given by Eqs.~(\ref{eq:ghzs}), (\ref{eq:ooxisb}), and quantum expectation,
\begin{eqnarray}
  \label{eq:jpfao}
  P_{x_A x_B x_C}(p,q,r) &=& \left|{}_{xxx}\langle p,q,r| \psi_{lmn} \rangle\right|^2 \nonumber \\
  &=& \frac{1}{d^2} ~\delta(p+q+r\equiv l \mod d),
\end{eqnarray}
where $\delta(k\equiv 0 \mod d)$ is defined to be unity if $k$ is congruent to zero modulo $d$ and zero otherwise. Summing $P_{x_Ax_Bx_C}(p,q,r)$ over Alice's outcome $p$ leads to the joint probabilities $P_{x_B x_C}(q,r)$ of Bob and Charlie obtaining outcomes $q$ and $r$:
\begin{eqnarray}
  \label{eq:jpfbc}
  P_{x_B x_C}(q,r) = \sum_{p=0}^{d-1} P_{x_Ax_Bx_C}(p,q,r) = \frac{1}{d^2}.
\end{eqnarray}
Note that the composite measurement $x_Bx_C$, when Bob and Charlie both measure in X, with Alice performing no measurement, results in the same joint probabilities as Eq.~(\ref{eq:jpfbc}), leaving no ambiguity in the expression ``$P_{x_B x_C}$.''

The conditional probabilities for the composite measurement $x_Ax_Bx_C$ are now obtained by substituting Eqs.~(\ref{eq:jpfao}) and (\ref{eq:jpfbc}) into Eq.~(\ref{eq:cpft}),
\begin{eqnarray}
\label{eq:c1}
P_{x_Ax_Bx_C}(p|q,r) &=& \delta(p+q+r \equiv l   \mod  d).
\end{eqnarray}
If $ p+q+r \equiv l \mod d$, $P_{x_Ax_Bx_C}(p|q,r) = 1$, implying perfect correlations among the pertaining outcomes $p$, $q$, and $r$ in $x_Ax_Bx_C$. Following similar procedures for the given GHZ state $|\psi_{lmn} \rangle$, we obtain conditional probabilities,
\begin{eqnarray}
\label{eq:c2}
P_{z_Az_B}(p|q) &=& \delta(p-q\equiv m   \mod  d),\\
\label{eq:c3}
P_{z_Az_C}(p|r) &=& \delta(p-r \equiv n   \mod  d),
\end{eqnarray}
for composite measurements $z_Az_B$ and $z_Az_C$, respectively.

The perfect correlations for a GHZ state $|\psi_{lmn}\rangle$ were shown in terms of the conditional probabilities in Eqs.~(\ref{eq:c1})-(\ref{eq:c3}). In an equivalent but alternative way, they can be characterized by showing that a set of secular equations satisfy with respect to some composite observable operators, called correlation operators. Consider three correlation operators, $X \otimes X\otimes X$, $Z \otimes Z^{\dagger} \otimes \openone$, and $Z \otimes \openone \otimes Z^{\dagger}$, with $\openone$ the identity operator. These will be denoted as $X_A X_B X_C$, $Z_AZ_{B}^{\dagger}$, and  $Z_AZ_C^{\dagger}$, respectively, for a sake of simplicity. The perfect correlations revealed by Eqs.~(\ref{eq:c1})-(\ref{eq:c3}) correspond to the following eigenvalue relations:
\begin{eqnarray}
  \label{eq:eeco1}
X_A X_B X_C |\psi_{lmn} \rangle &= \omega^l |\psi_{lmn} \rangle,\\
  \label{eq:eeco2}
Z_AZ_{B}^{\dagger} |\psi_{lmn} \rangle &= \omega^m |\psi_{lmn} \rangle,\\
  \label{eq:eeco3}
Z_AZ_C^{\dagger} |\psi_{lmn} \rangle &= \omega^n |\psi_{lmn} \rangle.
\end{eqnarray}
It is clear that the subscripts $lmn$ in GHZ states $|\psi_{lmn} \rangle$ stand for the eigenvalues, or strictly their exponents, for the three correlation operators. The characteristics of perfect correlations in Eqs.~(\ref{eq:c1})-(\ref{eq:c3}) or Eqs.~(\ref{eq:eeco1})-(\ref{eq:eeco3}) play a crucial role in detecting transmission errors caused by the interaction with a noisy environment.

\section{Detection of Transmission Errors}
\label{sec:dec}

When it is transmitted to a remote place, a system inevitably interacts with an environment. The interaction causes undesired changes in the system. These changes may even include losing the system. To simplify our error model we exclude such a possibility, which may be overcome by means of post-selection. The changes in a given state of the system are regarded as errors and are described by a trace-preserving completely positive quantum operation \cite{kraus}. Let $\rho$ be an initial state and $\rho'$ be a contaminated state when transmitted through a noisy environment. The contaminated state $\rho'$ can be expressed in Kraus representation by
\begin{equation}
\rho'=\sum_{\alpha} K_\alpha\rho K_\alpha^{\dagger},
\end{equation}
where Kraus operators satisfy the completeness relation, $\sum_\alpha K_\alpha^{\dagger}K_\alpha=\openone$. Each Kraus operator $K_\alpha$ stands for an error during the transmission. We will introduce an operator basis, known as an error basis, to represent Kraus operators by linear combinations in the error basis. The error basis is assumed for a qudit to be
\begin{equation}
\label{eq:eb}
E=\{\varepsilon_{i,j}=Z^i X^j |{i,j\in \{0,1,...,d-1\}}\},
\end{equation}
where unitary operators $X$ and $Z$ are defined in Eq.~(\ref{eq:xz}) and are called  level-shift and phase-shift errors respectively when used for representing errors. Note that the error basis $E$ forms an orthogonal basis on the Hilbert-Schmidt space of operators as $\mbox{Tr}(\varepsilon_{i,j}^\dagger \varepsilon_{k,l}) = d \delta_{i,k} \delta_{j,l}$ where $\delta_{i,k}$ is a Kronecker delta.

The correlation properties discussed in Sec.~\ref{sec:cor} play a crucial role in detecting transmission errors. To see this explicitly, consider the distribution of three qudits in a GHZ state $\ket{\psi_{000}}$ to three remote parties, Alice, Bob, and Charlie over a noisy environment. The initial state $\ket{\psi_{000}}$ is an eigenstate with the same eigenvalues $\omega^0=1$ for all the correlation operators, $X_AX_BX_C$, $Z_AZ_B^{\dagger}$, and $Z_AZ_C^{\dagger}$, as discussed in Sec.~\ref{sec:cor}. When errors occur during the transmission, the initial state $\ket{\psi_{000}}$ changes. If errors belong to $E$, they change it to another GHZ state in Eq.~(\ref{eq:ghzs}). It is obvious that the changed state can not have the same eigenvalues $\omega^0$ for all the correlation operators. Such an alteration can be utilized in order to detect transmission errors. For example, supposing that a level-shift error $X$ occurs on Alice's qudit, then the state changes to $X_A \ket{\psi_{000}}=\ket{\psi_{011}}$, which is an $\omega$-eigenstate of $Z_AZ_B^{\dagger}$ according to Eq.~(\ref{eq:eeco2}). If Alice and Bob measure $Z_A$ and $Z_{B}$ on the contaminated state $\ket{\psi_{011}}$, respectively, then their outcomes $p$ and $q$ satisfy $p-q=1$ as implied by Eq.~(\ref{eq:c2}). Due to the relation $p=q$ being satisfied if no error occurred, the error $X_A$ can be detected in the composite measurement $z_A z_B$. On the other hand, supposing that a phase-shift error $Z$ occurs on $B$, then the state changes to $Z_B \ket{\phi_{000}}=\ket{\phi_{100}} $, which is an $\omega$-eigenstate of $X_AX_BX_C$ as shown in Eq.~(\ref{eq:eeco1}). If Alice, Bob, and Charlie all measure $X$'s on their respective qudits, then their outcomes $p$, $q$, and $r$ satisfy $p+q+r=1$ according to Eq.~(\ref{eq:c1}). The error $Z_B$ can thus be detected in the measurement $x_Ax_Bx_C$ as $p+q+r=0$ should be satisfied if there was no error. In a similar way, any errors in $E$ can be detected by appropriate composite measurements. Note that we have so far described a direct measurement on a single copy of a GHZ state for error detection. On the other hand, composite measurements must collectively be performed on several copies of GHZ states in QECC's/QEDC's \cite{shor,gottesman}. This is also the case in our EPP, which will be presented in the following section.

We have assumed that the errors occurring belong to the error basis $E$, neglecting general errors of linear combinations in $E$. In this assumption, if an error in $E$ occurs, the initial GHZ state $|\psi_{000}\rangle$ changes to another GHZ state. Due to stochastic occurrence of errors, the contaminated state $\rho'$ is a statistical mixture of GHZ states. In other words, $\rho'$ is in a diagonal form in the basis of GHZ states, defined in Eq.~(\ref{eq:ghzs}). On the other hand, if more general errors are allowed, the contaminated state $\rho'$ may also include off-diagonal elements in the GHZ basis.  However, we use the measurements $X_AX_BX_C$, $Z_AZ_B$, and $Z_AZ_C$ for error detections in our EPP. As their basis are GHZ states (see Eqs.~(\ref{eq:eeco1})-(\ref{eq:eeco3})), their output states from $\rho'$ will be statistical mixtures of GHZ states even if $\rho'$ is not. Letting $\rho''$ be a purified state from the input $\rho'$, the diagonal elements of $\rho''$ depend only on those of $\rho'$. If purified successfully, a certain diagonal element of $\rho''$ is close to unity. The independence of the off-diagonal elements in $\rho'$ implies that neglecting general errors does not influence the performance of our EPP. The concise model of errors that belong to $E$ suffices for the error detections and significantly simplifies the analysis in our EPP. These kind of error models have also been employed in other EPP's \cite{bennett02, murao}.

\section{EPP for GHZ states}
\label{sec:epp}

Entanglement purification is concerned with obtaining a reduced number of copies of a high-fidelity entangled state from many polluted copies of noisy entangled states: It restores the quantum coherence of system polluted by environmental noise. In this section we propose an EPP for qudits which were initially in a GHZ state, say $|\psi_{000}\rangle$. As they were transmitted in a noisy environment, the qudits have been affected by the stochastic errors modeled in Sec.~\ref{sec:dec} and their state becomes a statistical mixture of GHZ states, given by
\begin{equation}
\label{eq:mghzs}
 \rho=\sum_{l,m,n=0}^{d-1}p_{lmn}\ket{\psi_{lmn}}\bra{\psi_{lmn}}.
\end{equation}
The copies that have transmission errors can be detected by an appropriate composite measurement due to the perfect correlations of GHZ states, as discussed in Sec.~\ref{sec:cor} and \ref{sec:dec}. Those copies are discarded by means of post-selection based on the measurement result. Repeating the whole process, the state of copies remaining is close to the error-free state $|\psi_{000} \rangle$, provided that the fidelity of $\rho$ to $|\psi_{000}\rangle$ is higher than a certain threshold value. The fidelity $p_{000}=\bra{\psi_{000}}\rho \ket{\psi_{000}}$ is used as a measure to investigate the performance of the EPP. If the fidelity $p_{000}$ converges to unity, then the purification is said to be successful.

\begin{figure}[!]
\includegraphics[width=0.4\textwidth]{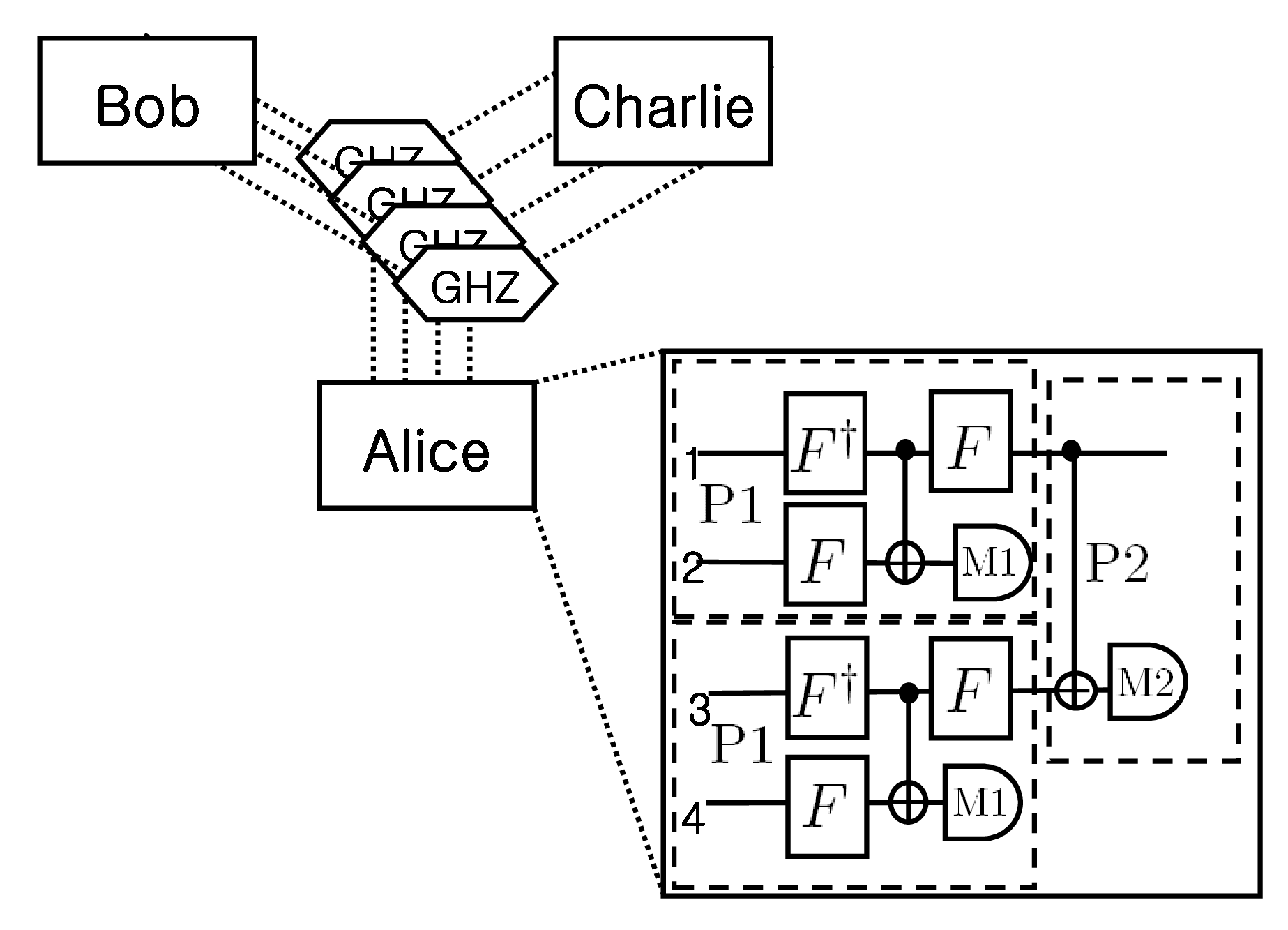}
\caption{A schematic configuration of the entanglement purification for three-qudit GHZ states. Alice, Bob, and Charlie share four noisy copies, labeled as 1,2,3 and 4. They perform local measurements and communicate via classical channels to detect any transmission errors.  Dashed box P1 (P2) is designed to detect and eliminate phase-shift (level-shift) errors.}
\label{fig1}
\end{figure}

The present EPP is described in more detail for three parties sharing qudits in the mixed state $\rho$ in Eq.~(\ref{eq:mghzs}). In this protocol every party performs sequentially two subroutines P1 and P2, that respectively correspond to the composite measurements $x_Ax_Bx_C$ and $z_Az_Bz_C$ for detecting phase-shift and level-shift errors. They are schematically shown by two dashed boxes P1 and P2 in Fig.~\ref{fig1}. P1 and P2 are designed to eliminate phase-shift and level-shift errors in the error basis $E$ (Eq.~(\ref{eq:eb})) and are performed sequentially. To eliminate phase-shift error, the three parties receive two copies of GHZ states and all perform the subroutine P1 on those. Alice, for instance, performs the inverse Fourier transformation $F^{\dagger}$ on qudit 1 from one copy, the Fourier transformation $F$ on qudit 2 from the other copy, and generalized XOR (GXOR) on them with qudits 1 and 2 as a control and target qudits, respectively. Here, GXOR is defined by
\begin{eqnarray}
\mbox{GXOR} |i\rangle_c |j\rangle_t = |i\rangle_c |i\ominus j\rangle_t,
\end{eqnarray}
where the subscript $c$ ($t$) indicates a control (target) qudit. Alice measures qudit 2 in the standard basis (defined below Eq.~(\ref{eq:ghzs})). The other parties do the same procedure on their qudits. After announcing their outcomes $p$, $q$, and $r$, the three parties keep the control qudits if $p+q+r \equiv 0 \mod d$ is satisfied. Otherwise, they discard those qudits. They repeat the previous process, until keeping two control copies. Applying the Fourier transformation $F$ to the two copies completes the subroutine P1. The subroutine P2 starts by applying GXOR to the two copies kept, for instance Alice applies GXOR to qudits 1 and 3. Alice measures the target qudit 3 in the standard basis, as in the box P2, Fig.~\ref{fig1}. The other parties do the same. All the parties announce their outcomes $p$, $q$, and $r$. If all the outcomes are the same, $p=q=r$, then they keep the control qudits, otherwise they discard them. The two post-selections will reduce the errors and increase the fidelity of the copies remaining. The whole process is repeated until a sufficiently high fidelity $p_{000}$ is obtained.

To analyze the performance of the present purification protocol, consider two copies in a composite state $|\psi_{lmn} \rangle \otimes |\psi_{l'm'n'}\rangle$ which the parties share. In the subroutine P1, just before the measurements, the composite state is transformed to
\begin{widetext}
 \begin{equation}
\sum_{p+q+r\equiv l}~~~\sum_{p'+q'+r'\equiv -l'} \omega^{(mq+nr-m'q'-n'r')} |p,q,r \rangle |p \ominus p',q \ominus q',r \ominus r' \rangle,
\end{equation}
\end{widetext}
where ``$p \equiv l$'' implies ``$p \equiv l \mod{d}$.'' After their measurements on the target qudits, the parties sum the outcomes and obtain $(p \ominus p')+(q\ominus q')+(r\ominus r')$. The sum is equal to $l+l' \mod{d}$, noting that $p+q+r\equiv l \mod{d}$ and $p'+q'+r'\equiv -l' \mod{d}$. If the sum $l+l' \equiv 0 \mod{d}$, the parties keep and perform $F$ on their control qudits, which are now in the state $|\psi_{l,m \ominus m',n \ominus n'} \rangle$. This complete the subroutine P1. In the subroutine P2, consider two copies in a composite state $|\psi_{lmn} \rangle \otimes |\psi_{l'm'n'}\rangle$ which are kept in P1. In their measurements on the target qudits following the GXOR operations, the parties will have all the same outcomes if $m=m'$ and $n=n'$, and keep the control qudits. In this case, these qudits are in the state $|\psi_{l \oplus l',m,n} \rangle$ where $l \oplus l'$ denotes $l+l' \mod d$.

When two input states in P1 are statistical mixtures $\rho = \sum p_{lmn} \ket{\psi_{lmn}}\bra{\psi_{lmn}}$ in Eq.~(\ref{eq:mghzs}), the output state is a new statistical mixture $\rho' = \sum_{l,m,n} p'_{lmn} \ket{\psi_{lmn}}\bra{\psi_{lmn}} $ with recurrence relations between $p_{lmn}$ and $p'_{lmn}$. One of these is given by
\begin{equation}
\label{eq:fpiepp}
p'_{000}=\frac{\sum_{m,n=0}^{d-1}p_{0mn}p_{0mn}}{\sum_{l,m,n,m',n'=0}^{d-1}p_{lmn}p_{d\ominus
l,m'n'}}.
\end{equation}
The output state in P2 for the given input $\rho'$ becomes another mixture $\rho''=\sum_{l,m,n}p''_{lmn}\ket{\psi_{lmn}}\bra{\psi_{lmn}}$. The recurrence relation between $p''_{000}$ and $p'_{lmn}$ is given by
\begin{equation}
\label{eq:spiepp}
p''_{000}=\frac{\sum_{l=0}^{d-1}p'_{l00}p'_{d\ominus
l,00}}{\sum_{l,l',m,n=0}^{d-1}p'_{lmn}p'_{l'mn}}
\end{equation}
If $p'_{000} > p_{000}$ and $p''_{000} > p'_{000}$, then we obtain a purified state through the subroutines P1 and P2. The conditions are reduced to a threshold fidelity such that if an initial fidelity is higher than it, the purification will succeed. With the help of the recurrence relations including Eqs.~(\ref{eq:fpiepp}) and (\ref{eq:spiepp}), we can numerically determine whether the present protocol successfully extracts $\ket{\psi_{000}}$ from statistical mixtures. The present protocol will be called a direct EPP.

We shall now present another protocol by generalizing the proposals of Ref.~\cite{horo} for high-dimensional Bell states of qudits. This protocol is an indirect EPP in the sense that GHZ states will be purified through the purification of Bell states in the way. It may be regarded as a simple extension from the bipartite EPP, as discussed by other authors \cite{murao}. The indirect protocol takes the following steps:

(1) An ensemble of qudits in noisy GHZ states, $\rho$ in Eq.~(\ref{eq:mghzs}), is divided into an equal amount of two subensembles.

(2) From each subensemble, Alice, Bob and Charlie share a trio of qudits in a noisy GHZ state. For the trio of qudits from one subensemble, Bob measures his qudit in the basis $\{ |k \rangle_x \}$ and announces the outcome $k$ to the others. Receiving the outcome $k$, Alice applies an operation $Z^k$ to her qudit, where $Z^k$ is the $k$-th power of $Z$ in Eq.~(\ref{eq:xz}). Then, Alice and Charlie have the pair of qudits in a reduced (high-dimensional) Bell state. From the other subensemble, Charlie and Alice do the same as Bob and Alice did, and Alice and Bob have another pair in a reduced Bell state. This step is repeated over the subensembles. Alice and Bob have one subensemble of qudits in noisy Bell states $\rho_{1}$ and Alice and Charlie have the other in $\rho_{2}$.

(3) Alice and Bob (Alice and Charlie) purify their noisy Bell states, $\rho_1$ ($\rho_2$), by performing the EPP by Alber {\em et. al.} \cite{horo}. They obtain a reduced number of copies in purified Bell states, $\rho'_{1}$ ($\rho'_{2}$), if $\rho$ has a fidelity higher than a certain threshold.

(4) From each post-selected subensemble, Alice chooses a pair of qudits. She then share one with Bob and the other with Charlie. She performs a GXOR operation on her qudits, where a control (target) qudit belongs to the pair shared with Bob (Charlie). Measuring the target qudit in the standard basis $\{ |k\rangle\}$, she announces the outcome $k$ and Charlie applies $X^{d \ominus k}$ to his qudit. They obtain then a state closer to $\ket{\psi_{000}}$. This step is repeated over the post-selected ensembles.

\begin{table}[!]
\caption{Threshold fidelities $f_\mathrm{dir}$ and $f_\mathrm{ind}$, the minimum fidelities required for the successful purification by the direct and indirect EPP's are presented for noisy GHZ states, as given in Eq.~(\ref{eq:cdiepp}). The number $d$ is a dimension of the Hilbert space for each qudit.}
\begin{ruledtabular}
\begin{tabular}{ccc}
$d$ & $f_\mathrm{dir}$ & $f_\mathrm{ind}$ \\ \hline
 2 & 0.4073 & 0.4167 \\
 3 & 0.2305 & 0.3000 \\
 4 & 0.1555 & 0.2227 \\
 5 & 0.1155 & 0.1780 \\
 6 & 0.0907 & 0.1489 \\
\end{tabular}
\end{ruledtabular}
\label{table1}
\end{table}

We compare the direct and indirect EPP's in terms of their threshold fidelities, $f_\mathrm{dir}$ and $f_\mathrm{ind}$ respectively, for successful purification and the minimum numbers of copies required to achieve a certain high fidelity, say 0.99. For the comparison, consider a mixed state of
\begin{equation}
\label{eq:cdiepp}
  x|\psi_{000} \rangle
\langle \psi_{000}| + \frac{(1-x)}{d^3} \openone,
\end{equation}
where $\openone/d^3$ is a purely random state of three qudits. In Table~\ref{table1} we present the threshold fidelities $f_\mathrm{dir}$ and $f_\mathrm{ind}$ in terms of dimension $d$. The thresholds decrease as the dimension $d$ increases, implying that both EPP's work for lower fidelities of mixtures at higher dimensions. For each dimension $d$, the direct protocol has a lower threshold than its counterpart and in a way the direct EPP can be seen to work better than the indirect. The direct protocol is also more efficient in the sense that it requires a smaller number of copies to achieve a certain high fidelity. For instance, considering a case of $d=6$, suppose that the initial state in Eq.~(\ref{eq:cdiepp}) is of fidelity 0.5. In the direct protocol, about 48 copies suffice to achieve fidelity $0.99$. On the other hand, the indirect EPP needs about 192 copies. These comparisons in terms of the threshold fidelities and the minimum numbers of copies leads us to conclude that the direct protocol shows better performance than its counterpart. It is because the direct protocol utilizes multipartite correlations, whereas the indirect one employs only bipartite correlations. Utilizing multipartite correlations leads to efficient entanglement purification.

\section{Remarks}
\label{sec:rem}

With the present direct EPP we perform alternatively the subroutines P1 and P2. Other combinations of subroutines, for instance P1+P1+P2 are potentially possible. Such variants may be beneficial against some transmission errors. For instance, it was argued in Ref.~\cite{thew} that phase errors are dominant in experiments for high-dimensional entangled states. In this case the subroutine P1 could be employed more frequently than P2. A detailed analysis of such variants is beyond the scope of this paper.

A purification protocol utilizes a detecting scheme of errors rather than a correcting scheme. However, the results from QECC's are also important as they can easily be adopted to QEDC's. QECC's, in particular Calderbank-Shor-Steane codes treat separately bit-flip and phase-flip errors \cite{shor}. This treatment of separating errors was shown to be useful in developing EPP's for qubits \cite{chen}, which have in common two subroutines similar to our P1 and P2.



In summary, we have investigated and analyzed the correlation properties of GHZ states for qudits ($d$-dimensional systems). It was shown that a noisy environment alters the correlations of a GHZ state if it contaminates the state of qudits during the transmission. Based on this, we have suggested a detection scheme for errors and then a direct EPP that purifies noisy GHZ states of qudits. By comparing with the indirect EPP derived from the EPP's for two-qudit Bell states, we showed that the direct EPP shows better performance than the indirect one in terms of threshold fidelities and the minimum numbers of copies for achieving a high fidelity.

\begin{acknowledgments}

We thank Dr. M. Tame for reading and commenting an earlier version of this paper. This work was supported by MOST/KOSEF through the Quantum Photonic Science Research Center, the Leading Basic S\&T Research Projects, and by the Korean Research Foundation Grant funded by the Korean Government (MOEHRD) (KRF-2005-041-C00197). S.-W. Lee was supported by the EU through the STREP project OLAQUI.

\end{acknowledgments}

\end{document}